\documentclass[12pt,preprint]{aastex}
\begin{document}


\def\pdot {\dot P}
\def\xte{\textit{RossiXTE}}
\def\asca{\textit{ASCA}}
\def\sax{\textit{BeppoSAX}}
\def\psr{PSR~J1846--0258}
\def\kes{Kes~75}

\title{\sax\ observations of the young pulsar in the \kes\ Supernova Remnant}
\author{S. Mereghetti}
\affil{Istituto di Astrofisica Spaziale e Fisica Cosmica -
Sezione di Milano ``G.Occhialini'' - CNR \\
v.Bassini 15, I-20133 Milano, Italy \\
sandro@ifctr.mi.cnr.it}
\author{R. Bandiera}
\affil{Osservatorio Astrofisico di Arcetri   \\
Largo E. Fermi 5, I-50125 Firenze, Italy \\
bandiera@arcetri.astro.it}
\author{F. Bocchino}
\affil{Osservatorio Astronomico di Palermo    \\
Piazza del Parlamento 1, I-90134 Palermo, Italy \\
bocchino@astropa.unipa.it}
\author{G.L. Israel}
\affil{Osservatorio Astronomico di Roma    \\
v.Frascati 33, I-00040 Monteporzio  Catone, Italy \\
gianluca@mporzio.astro.it}

\begin{abstract}

We present the results of  \sax\  observations of the young
X--ray pulsar \psr, recently discovered at the center of the
composite supernova remnant \kes. The pulsar (plus nebula) spectrum can be fitted by
an absorbed power law with photon index
$\alpha_{\rm ph}=2.16\pm0.15$,
$N_{\rm H}=(4.7\pm0.8)\times10^{22}$~cm$^{-2}$, and
unabsorbed flux $\sim\!3.9\times10^{-11}$~erg~cm$^{-2}$~s$^{-1}$ (2--10~keV).
By joining two observations taken at an interval of two weeks we have been able
to obtain a precise measurement of the spin period ($P=324.818968\pm0.000006$~ms).
This value, when combined with previous measurements, cannot be fitted by
a smooth frequency evolution with a canonical braking index $n=3$.
With the hypothesis of no glitches and/or significant timing noise,
the braking index would be  $n=1.86\pm0.08$
and, assuming a short initial period, the pulsar age
would be $\sim$1700 years, closer to that    of the   supernova remnant
than the simple estimate   $\tau=P/2\pdot=723$~years.
Other likely possibilities involve the presence of glitches
and lead to a wide range of acceptable ages.
For example, we obtain $n$ in the range 1.8--2.5, if a  glitch occurred near MJD 51500
while for a glitch between October 1993 and  March 1999 we can only get a lower limit
of $n>1.89$.

\end{abstract}


\section{Introduction}

The X--ray pulsar \psr\ ($P=325$~ms) was discovered by Gotthelf et al.\ (2000)
at the center of the composite supernova remnant \kes\ (G29.7--0.3).
Up to now, no radio detection of \psr\ has been reported, with an
upper limit of $\sim\!0.1$~mJy at 1.5~GHz (Kaspi et al.\ 1996).
All the information  on the pulsar timing parameters has been determined through
observations in the X--ray band. Gotthelf et al.\ (2000) obtained
a period derivative of $\pdot=7.1\times10^{-12}$~s~s$^{-1}$, by
comparing five period measurements obtained with
the \xte\ and \asca\ satellites in the years from 1993 to 2000.
Assuming the canonical relation for the spin-down by magnetic dipole
radiation, these values lead to an estimate for the magnetic field
of $B\!\sim\!5\times10^{13}$~G,
above the quantum critical field and in the range
of the highest values observed in radio pulsars.
Even more remarkably, the  characteristic age of \psr,
$\tau=P/2\pdot=723$~years,
is the smallest one of any known pulsar.

The association of \psr\ with the supernova remnant \kes\ can be regarded as
almost certain: the pulsar is located at the geometrical center
of the $3.5'$ diameter shell (Collins, Gotthelf \& Helfand 2002)
and is powering a bright radio/X--ray nebula that gives the composite morphology to this
supernova remnant.
The estimate of the age of \kes\ is subject to a large uncertainty.
Assuming that the shell is still in the freely expanding phase with
velocity $v$, the age is $1800\,d_{19}\,(v/5000\,{\rm km\,s^{-1}})^{-1}$~years,
where $d_{19}$ is the distance normalized to the value of 19~kpc,
estimated for \kes\ with 21~cm observations (Becker \& Helfand 1984).
Alternatively, if the remnant is already in the Sedov phase, the age
can be estimated by the relation between the radius and the shock temperature
inferred from the X--ray spectra. In this way Blanton \& Helfand (1996) derived
an age of $7000\,d_{19}$~years, based on the temperature of 0.5~keV measured
with \asca.
It thus appears that   the pulsar characteristic age
is smaller than the age of \kes,
although the large uncertainties
involved do not allow agreement between the two values to be
precluded.

\section{Observations and Data Analysis}

The location of \psr\ was observed twice with the \sax\ satellite in March 2001
(see Table~1).
The instrument relevant for the observations reported here is
the   Medium-Energy Concentrator
Spectrometer (MECS), that    operates in  the 1.8--10~keV energy range,
providing a good spatial ($\sim\!1'$ full width at half maximum (FWHM))
and energy resolution ($\sim\!8.5$\%  FWHM at 6~keV)  over a circular
field of view with a diameter of $56'$ (Boella et al.\ 1997).

The main target of the first observation (A)
was  the   pulsar AX~J1844--0258, therefore \psr\ was detected at an off-axis
angle of $\sim\!23'$  (results on AX~J1844--0258 will be reported elsewhere).
The second observation (B) was  pointed on the Kes 75 supernova remnant.
Both observations lasted about 2.5~days, but
Earth occultations and passage of the satellite in regions
of high particle background resulted in net exposure times of 83.7~ks
and 105.3~ks for the MECS instrument.

\subsection{Timing analysis}

For the timing analysis we used only counts with energy greater than
3~keV in order to reduce the contribution of  the soft emission from the
supernova remnant. For observation B  we used a circular extraction
region with radius $2'$ (resulting in 15904 counts),
while for observation A we used an elliptical
region matched to the shape of the off-axis MECS point spread function (3751 counts).
The times of arrival were converted to the Solar System barycenter
using the source position
R.A.$=18^{\rm h}\,46^{\rm m}\,24.5^{\rm s}$,
Dec.$=-02^{\circ}\,58'\,28''$ (J2000).

We first analyzed the two observations separately, using a folding program
to search over a grid of period ($P$) and period derivative ($\pdot$) values
and adopting the method
described in Leahy (1987) to determine the best period and its uncertainty.
Since the individual observations are relatively short compared to the
pulsar spin-down timescale, the period values giving
the highest signal do not depend significantly on $\pdot$. We therefore fixed
$\pdot$ to the value $7.097\times10^{-12}$~s~s$^{-1}$
 (Gotthelf et al.\ 2000)
obtaining the period values reported in Table~1.
The light curve obtained from observation B is shown, for different
energy ranges, in Fig.~1.
The difference between the periods measured in the two observations corresponds to
$\pdot=(7.127\pm0.096)\times10^{-12}$~s~s$^{-1}$ which is,
within the errors, compatible with the value found
from the \asca\ and \xte\ data spanning the years 1993--2000
(Gotthelf et al.\ 2000).

The accuracy of the \sax\ satellite clock is good enough to
combine the two observations (taken about 15 days apart)
into a single data set, thus allowing us to derive a smaller error on
$P$ and to directly measure $\pdot$.
This was again done by folding the data on a grid of  $P$ and $\pdot$ trial values,
leading to $P=324.818968\pm0.000006$~ms and
$\pdot=(7.095\pm0.086)\times10^{-12}$~s~s$^{-1}$,
where the values refer to the epoch MJD=51991.08778.
These results are consistent with an extrapolation of the timing solution
derived by   Gotthelf et al.\ (2000).

\subsection{Spectral analysis}

For the spectral analysis  we used only observation B that provided
data of better quality since \psr\ was detected on-axis, where the MECS
instrument has the greatest sensitivity and the best spectral and
spatial performances.  Moreover, we used also data from the  LECS
instrument (Parmar et al. 1997) that covers the soft energy range down
to 0.1 keV.

For the source spectrum extraction we used first a circle of radius
4$'$ for the MECS and 8$'$ for the LECS in order to maximize the
contribution from both the supernova remnant and the pulsar nebula
emission.
The background spectrum was estimated from a circular
corona between 4.7$'$ and 9.5$'$  for the MECS and between 8$'$ and 10.6$'$ for
the LECS. The source spectra were rebinned in order to have at least 30
counts per channel.

We   verified that a fit with a power-law only fails
to describe the data ($\chi^2$/d.o.f. = 463/404), giving strong residuals
corresponding to the presence of the emission lines of Si XIII (1.84
keV) and S XV (2.45 keV).
We therefore fitted the spectra using a combination of a power law and an optically
thin thermal plasma ({\sc mekal} in the XSPEC softare package v.11),
plus an interstellar absorption of Morrison and McCammon (1983). This
combination of models is known to be appropriate for Kes 75, because
it also takes into account the presence of a thermal shell (Blanton \&
Helfand 1996).
This model produced a  very good fit ($\chi^2$/d.o.f. = 398/402)
with the following parameters: interstellar absorption
$N_H=4.7(3.9-5.5)\times 10^{22}$ cm$^{-2}$, photon index
$\alpha_{\rm ph} = 2.19 (2.07-2.30)$, plasma temperature $kT = 0.40(0.35-0.53)$ keV.
This best-fit is shown in Fig.\ 2. The thermal
and non-thermal fluxes in the 0.1--2 keV range  are respectively
$10.1(3.0-24.4)\times 10^{-10}$
and $9.9(7.9-12.0)\times 10^{-11}$ erg cm$^{-2}$ s$^{-1}$,
while the corresponding values in the 2--10 keV band
are $5.4(1.6-13.0)\times10^{-12}$ and
$3.4(2.7-4.1)\times 10^{-11}$ erg cm$^{-2}$ s$^{-1}$.
The contribution of the thermal component to the flux density drops below
10\% of the total above 3 keV. To further constrain the spectrum of the
nebula, we have repeated the analysis using the data collected from
a   smaller circle (2$'$ radius), and fixing the temperature of the
thermal component and the interstellar absorption to the values obtained previously.
This resulted in  a best-fit with $\alpha_{\rm ph} =
2.16(2.10-2.21)$  ($\chi^2$/d.o.f.=348/317). The non-thermal unabsorbed
flux in the 2--10 keV band is $3.9(3.5-4.2)\times 10^{-11}$ erg
cm$^{-2}$ s$^{-1}$.
These results are consistent with (and have smaller uncertainties than)
those obtained by Blanton \& Helfand (1996)
with \asca\, that, similarly to our instrument, did not have enough spatial resolution
to disentangle the contribution from the shell and the central pulsar/nebula.

We also examined the pulsar spectrum in two different phase intervals
corresponding to the high and low parts of the pulse profile, finding
marginal evidence for a hardening  in correspondence of
pulse maximum.  This indicates that the surrounding diffuse X--ray
nebula, giving a higher relative contribution during the phase of
low pulsed emission,  has a  spectrum softer than that of the pulsar.
In fact, the preliminary \textit{Chandra} results reported by Collins et al. (2002),
yield a power law photon index of $\sim$1.5 for the pulsar and $\sim$1.9 for the
nebula.

\section{Discussion}

The period measurement derived with \sax, together with past
measurements (as from Table~1 in Gotthelf et al.\ 2000),
allows an analysis of the
rotation history of \psr\ in the past decade. All together, there are
now 6 different measurements, of which ours is that with the smallest
uncertainty (just 12\% of the average of all the others).
In the case of a smooth evolution represented
with a quadratic function, of the form

$$\nu(t)=\nu_0+\dot\nu_0(t-t_0)+{1\over2}\ddot\nu_0(t-t_0)^2,$$

\noindent
a reasonable representation of the data can be achieved
with the best fit parameters reported in the first column of
Table~2 ($\chi^2=3.52$, for 3
d.o.f.; residuals are plotted in Fig.~3.b).
The best fit second derivative of the frequency $\ddot\nu_0$ is
significantly different from 0.
In fact a linear fit ($\ddot\nu_0=0$) can be safely ruled out, as shown
by the
extremely high $\chi^2$ value ($\sim\!534$, for 4 d.o.f.).
We also note that the residuals of a linear fit
(Fig.~3.a) are too large to be accounted for by timing noise,
even assuming for \psr\ a particularly high activity parameter (see, e.g.,
Arzoumanian et al. 1994).

Under the hypothesis that no glitch has occurred over the time
spanned by the period measurements, the available data can be
used to determine the braking index $n$, defined by the   law
$\dot\nu\propto-\nu^n$ and evaluated, in terms of the frequency and its first
two derivatives, as $n=\nu\ddot\nu/\dot\nu^2$.
The values of the quadratic fit  correspond to
a braking index $n=1.86\pm0.08$,
significantly lower than the
canonical value ($n=3$) for magnetic dipole losses. For a braking law
$\dot\nu\propto-\nu^n$, the pulsar age is:

$$(1)~~~~~~~~~~~~\tau={P\over(n-1)\dot P}\left[1-\left(P_0\over P\right)^{n-1}\right].$$

Assuming an originally fast spinning pulsar ($P_0\ll P$), we derive for \psr\
an age $\tau=1675\pm157$~yr.
The quoted accuracy on $\tau$ relies on the assumption that irregularities
in the spin evolution are negligible in the range of epochs spanned by the
observations. In particular, if found that one or more glitches have occurred in
between period measurements, the above estimates should be revised.
Although population studies indicate small initial periods for radio pulsars
(Bhattacharya et al. 1992,  Lorimer et al. 1993) and a value of
$P_0\sim17-19$~ms is generally accepted for the Crab pulsar,
there is now considerable evidence that fairly long
initial spin periods may be common
(see, e.g.,  Torii et al. 1999, Kaspi et al. 2001, Murray et al. 2002).
We therefore  show in Fig.~4 (solid line), how the age derived from
equation (1) depends on the assumed value of $P_0$.

The data can also be interpreted by considering the possibility that glitches
occurred between the observations.
For instance, let us first consider the case of a
glitch between the first (October 1993) and the second (March 1999) measurements
by comparing the October 1993 value with the extrapolation of a fit to the
subsequent measurements.
A quadratic fit to the   last five points
gives a poorly constrained  $\ddot\nu_0$  (corresponding to
$n=0.84\pm0.68$; see column (2) in Table~2).
The estimated magnitude and even the sign of the   glitch actually
depend on the   value of the braking index.
The best fit  ($n=0.84$)    corresponds to   a negative jump in
frequency, a behavior opposite to what seen in all glitching radio pulsars
(Lyne et al.\ 2000). A conventional sign is obtained only for $n>1.89$,
which being at only $\sim1.5\sigma$ from the best fit, has a non negligible probability.

Another possibility is the occurrence of a glitch at a later time.
In fact the residuals in Fig.~3.b
show a pattern suggesting  a glitch
in an epoch between the third and fourth data point. We have then chosen
51500.0~MJD as a possible epoch of this glitch: any different choice of the
epoch, within the time interval between the third and fourth data point,
will have the only effect of slightly changing the best parameters for the
glitch, therefore not changing any of the conclusions that we shall draw below.

For different values of the braking index, we have fitted separately the
former and latter set of three data points. We find that for values of $n$
larger than about 2.5 we would again come out with an unlikely ``anti-glitch''.
On the other hand, for $n$ smaller than about 1.8 the
(negative) $\dot\nu$ should increase, and also this behavior is in contrast
with what seen in radio pulsars (Lyne et al.\ 2000). In the allowed range
for $n$ (1.8,2.5), the $\chi^2$ value is monotonically decreasing for
increasing $n$, but the trend is so shallow that by itself it cannot be
taken as a valid argument for preferring higher values of $n$. The parameters
for the two limit cases are shown in columns (3) and (4) of Table~2,
and the corresponding pulsar ages by the dashed lines of Fig.~4.

If a glitch at about MJD 51500 actually occurred, how does it compare with
known glitches in other pulsars? Fig.~5 shows the positions, in the parameter
plane $(\Delta\nu/\nu)$--$(\Delta\dot\nu/\dot\nu)$ of all glitches known in
radio pulsars  for which both $\Delta\nu$ and
$\Delta\dot\nu$ have been measured (as from Lyne et al.\ 2000).
Superposed is the locus of the points
representing the MJD 51500 glitch, for different values of $n$. It is
apparent that the required glitch has rather normal parameters, with both
$(\Delta\nu/\nu)$ and $(\Delta\dot\nu/\dot\nu)$ somehow smaller than the
average.

\section{Conclusions}

We have presented a new accurate measurement of the spin frequency of the
young pulsar \psr\ associated to the \kes\ supernova remnant.
Combined with previous data, and making some hypothesis on the glitch history of
\psr\, our new frequency measurement can be interpreted in different ways.

A simple possibility, assuming no glitches
in the years 1993--2001, is in terms of a timing solution with
$\ddot{\nu}=(2.77\pm0.12)\times10^{-21}$~Hz~s$^{-2}$.
This    results in a braking index $n=1.86\pm0.08$ and
an estimate for the age of \psr\ more in
accord with that of the associated supernova remnant (unless
the pulsar initial period was close to the current one).
Other likely possibilities involve a single glitch with reasonable parameters
and lead to a wide range of acceptable ages.
If the glitch occurred near MJD 51500 the resulting braking index  $n$ is
in the range 1.8--2.5,
while for a glitch between October 1993 and  March 1999 we can only get a lower limit
of $n>1.89$.

Braking indices have so far been measured only for five pulsars.
Except for PSR J1119--67 which has $n=2.91\pm0.05$ (Camilo et al.\ 2000),
all of them  have $n$ significantly smaller than 3, similar to
the values suggested by our analysis for \psr. Such  values are
$2.51\pm0.01$ (Crab pulsar; Lyne, Pritchard \& Graham-Smith 1993),
$2.837\pm0.001$ (PSR B1509--58; Kaspi et al. 1994),
$1.4\pm0.2$ (Vela pulsar; Lyne et al. 1996), and
$1.81\pm0.07$ (PSR B0540--69; Zhang et al. 2001).

The pulsar with the highest glitch rate (about one per year) in the
sample studied by Lyne et al. (2000), is  PSR B1737--30, which is also the one with
the highest magnetic field ($1.7\times10^{13}$~G).
We cannot exclude that \psr\ had  more than one glitch in the eight years spanned by the
observations, but in this case no information on $n$ can
be derived from our data.
Interestingly, the glitch parameters we inferred for \psr\ in Fig.~5
are similar to those observed in PSR B1737--30.
A finer monitoring in the future will establish whether \psr\ is indeed
similar to PSR B1737--30.

As already mentioned by Gotthelf et al.\ (2000), the
$P$ and $\pdot$ values of \psr\ are very similar to those of the radio
pulsar PSR~J1119--6127
($P=407.6$~ms, $\pdot=4\times10^{-12}$~s~s$^{-1}$, Camilo et al.\ 2000).
A remarkable difference between these two young and energetic pulsars
($\dot E_{\rm rot} \gtrsim2\times10^{36}$~erg~s$^{-1}$) is the lack of a bright
pulsar wind nebula around PSR J1119--6127 (Crawford et al.\ 2001).
These authors suggested that high magnetic field pulsars produce
radio nebulae that fade rapidly, an interpretation apparently contradicted by
the brightness of the \kes\ core.
If the value $n=1.86$ is confirmed for \psr, the difference in the
brightness of the wind nebulae of these two pulsars might be related to their
different braking index values
($n=2.91\pm0.05$  for PSR J1119--6127; Camilo et al.\ (2000)).
While spin-down
by magnetic dipole radiation yields $n=3$, a value $n=2$ is
expected if  the pulsar braking is driven by the presence of a  strong relativistic
wind, which could also have an effect on the emission from a  surrounding nebula.

\acknowledgments
We thank C. Paizis for useful discussions.
This work has been partly supported by the Italian Ministry for University and
Research (MIUR) under Grant Cofin2001--02--10.

\begin{deluxetable}{cccccc}
\tablecaption{Log of \sax\ observations}
\tablehead{
\colhead{Observation}               &
\colhead{Start date} &
\colhead{MJD$^a$} &
\colhead{Exposure time} &
\colhead{Period$^b$} \\
 (1) & (2) & (3) & (4) & (5)
}
\startdata
 A & March 14, 2001   & 51983.27098   &  83716~s  & 324.81411(12)~ms   \nl
 B & March 29, 2001   & 51998.86160   &  105327~s & 324.82371(5)~ms   \nl
\hline
\enddata

\vspace{0.5cm}

$^a$ Modified Julian Date of the middle of the observation

$^b$ values referred to the Epoch of Column (3) and for fixed
     $\pdot=7.097\times10^{-12}$~s~s$^{-1}$ (as from Gotthelf et al.\ 2000)

\end{deluxetable}


 \begin{deluxetable}{ccccc}
 \tablecaption{Derived timing parameters for \psr\ (1$\sigma$ uncertainties)}
 \tablehead{
  \colhead{ }               &
 \colhead{all the points }               &
 \colhead{last five points }               &
 \colhead{all the points$^a$}               &
 \colhead{all the points$^a$} \\
 & (1) & (2) & (3) & (4)
 }
 \startdata
%
$t_o$ (MJD)        & 50000.0 & 50000.0 & 51500.0 & 51500.0 \nl
 & & & & \nl
 $\nu_0$  (Hz)         & 3.090255(1)   & 3.09024(1)      & 3.0814965(7)   &     3.0814947(7)  \nl
                       & &                                &   3.0814965(5)   &    3.0814963(5)  \nl
   &   & &  &    \nl
$\dot \nu_0$ ($10^{-11}$~Hz~s$^{-1}$)&--6.7768(14)&--6.754(15)&--6.7319(11)  &   --6.7422(11)  \nl
                                        & &       &  --6.7448(12)     &   --6.7422(12)  \nl
  &  & &   &      \nl
$\ddot \nu_0$ ($10^{-21}$~Hz~s$^{-2}$) & $2.77\pm0.12$ & $1.2\pm1.0$  &  [3.647]$^b$       &      [2.685]  \nl
                                                                 & &   & [3.661]      &       [2.685]  \nl
  &   & &      &      \nl
 $\chi^2$/d.o.f.                & 3.52/3  & 1.18/2 &    0.029/1            &  0.086/1  \nl
                        & &                & 0.240/1             & 0.508/1  \nl
 Total:                 & &             &    0.269/2           &  0.594/2  \nl
   &     & &    &     \nl
 $n$          & $1.86\pm0.08$  & $0.84\pm0.68$ &     2.48$^c$        &  1.82$^c$   \nl
   &        &   & &   \nl
$\tau$ (yr)         & $1675\pm157$ &                    & [980.1]       &    [1766.2]  \nl
                        & &              &    [978.2]        &   [1766.2]  \nl
   &    &    & &   \nl
 \enddata

  \vspace{0.5cm}

 $^a$  The upper and lower values in each line refer
 respectively to the fit to data points before and after MJD 51500.

 $^b$ All the values within square brackets strongly depend on the assumed $n$ value.

 $^c$ Assumed  value

 \end{deluxetable}

\clearpage

\figcaption{Folded light curve of \psr\ in three energy ranges,
obtained with the MECS instrument in the observation of March 29$^{\rm th}$,
2001.
\label{fig1}}

\figcaption{Best fit spectrum of \psr\ (upper panel) and residuals (lower panel).
The  points correspond to the LECS (0.5-9 keV) and MECS (1.8-10 keV) data.
The lines in the upper panel are the best fit model folded through the
instrumental response.
\label{fig2}}

\figcaption{Residuals of the fits to the spin frequency as a function of
time for the cases (a) of a linear   and (b) of a quadratic relation.
Note the different scales of the vertical axis.
The vertical line in plot (b) indicates the possible location of a glitch, that
may account for the pattern of the residuals (see text).
\label{fig3}}

\figcaption{Age of \psr\ as a function of its initial spin period. The solid line
refers to $n=1.86$, the dotted line to  $n=3$,
and the dashed lines indicate the range allowed by values
of $n$ in the interval 1.82 -- 2.48.
\label{fig4}}

\figcaption{Parameters of the glitch fitted to the \psr\ data compared with
the values of $(\Delta\nu/\nu)$ and $(\Delta\dot\nu/\dot\nu)$ of glitching pulsars
(Lyne et al.\ 2000).
The line indicates the values obtained for different
assumptions on the braking index value; the empty dots represent all braking
indices from 1.85 to 2.45, separated by 0.05.
The circled dots represent the glitches of the highest field radio pulsar in the
sample (PSR~1737--30, with $B=1.7\times10^{13}$~G): they have parameters
similar to those expected for the possible MJD~51500 glitch in \psr.
\label{fig5}}


\clearpage
\epsscale{.7}
\plotone{f1.eps}

\clearpage
\epsscale{.8}
\plotone{f2.eps}

\clearpage
\epsscale{.8}
\plotone{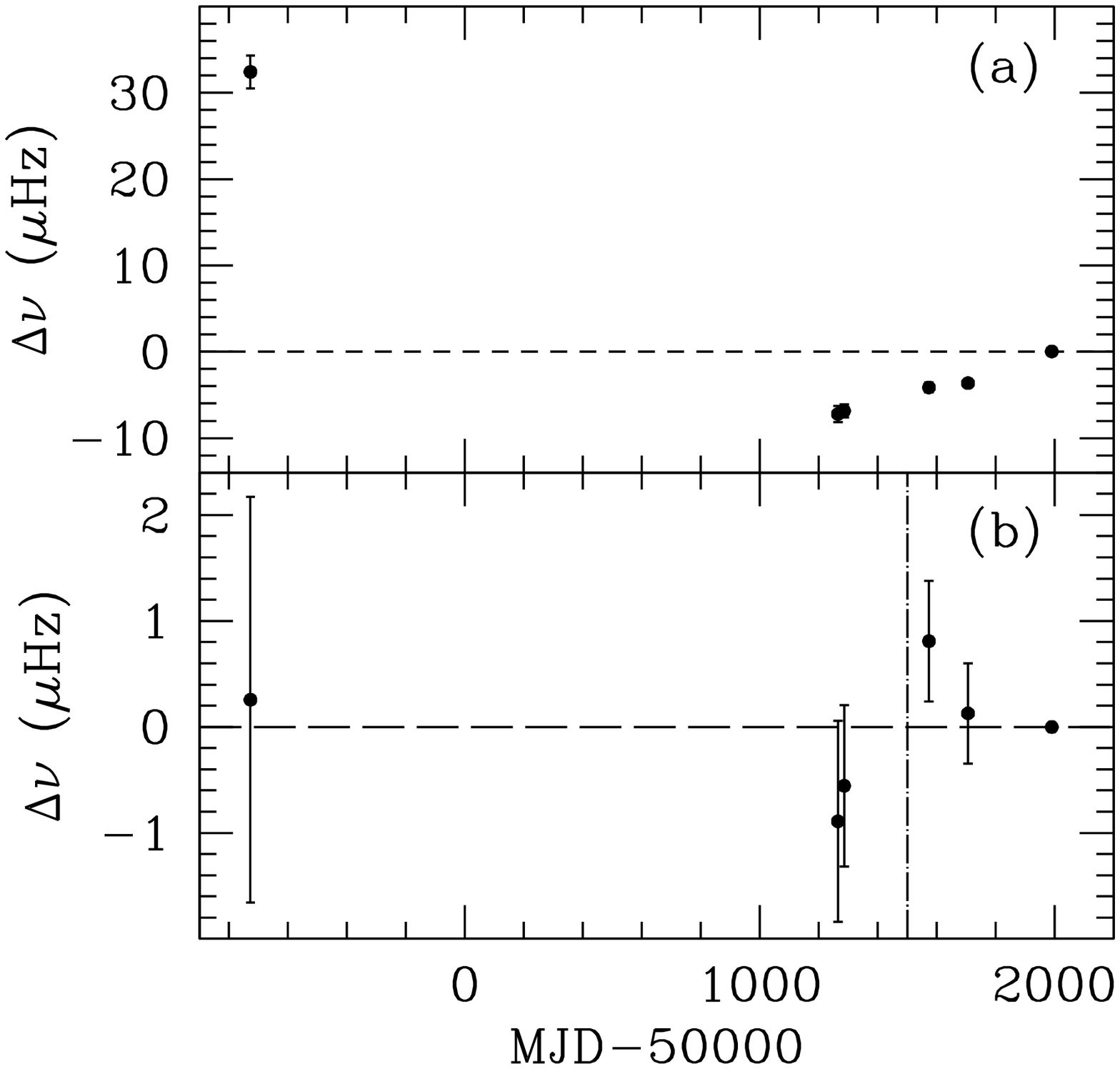}

\clearpage
\epsscale{.8}
\plotone{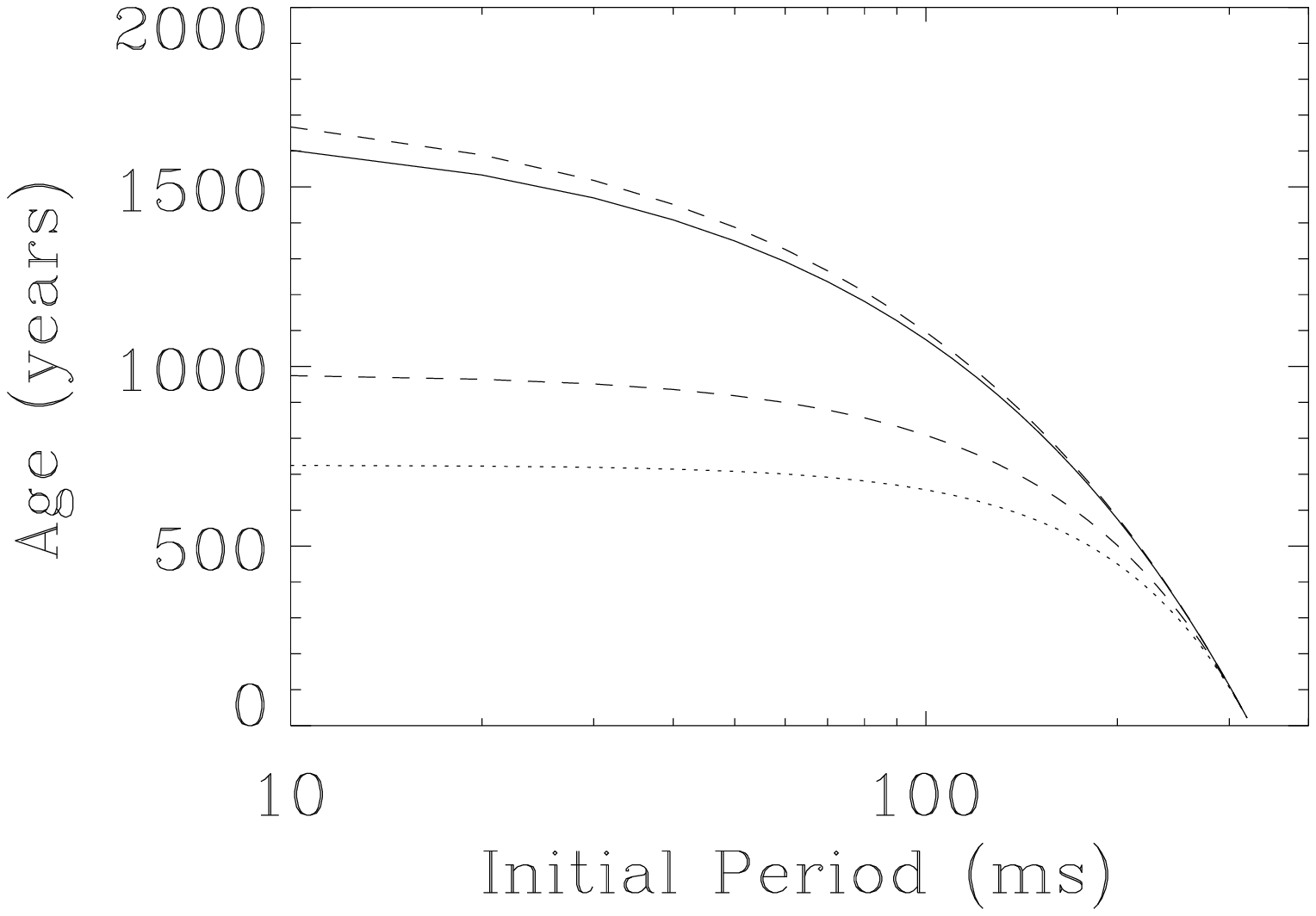}

\clearpage
\epsscale{.8}
\plotone{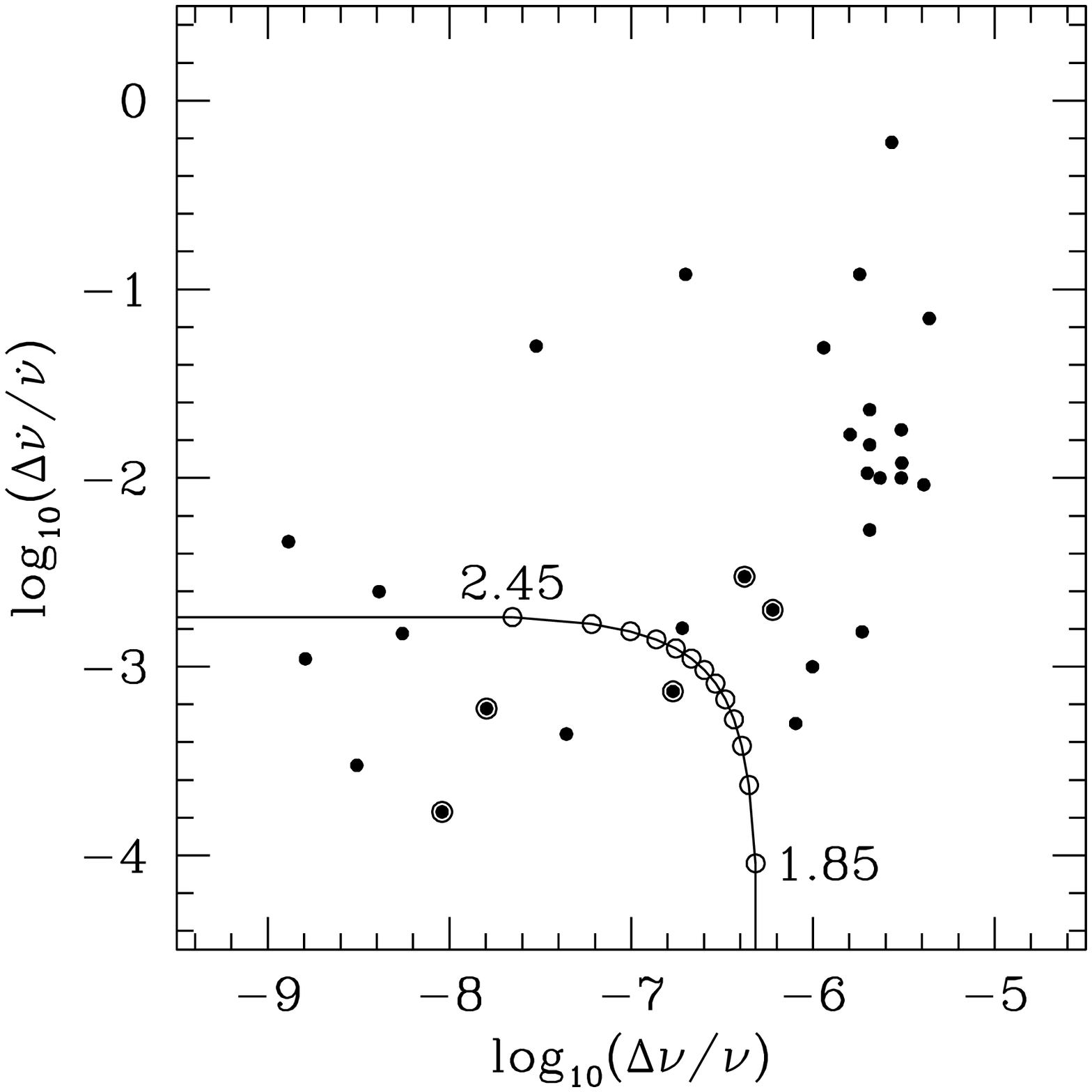}

\end{document}